\documentstyle[epsfig]{aipproc}

\def\degs{\ifmmode ^{\circ}\else$^{\circ}$\fi}

\def\kev{\hbox{ke\hskip -0.16ex V}}

\newbox\grsign \setbox\grsign=\hbox{$>$}
\newdimen\grdimen \grdimen=\ht\grsign
\newbox\laxbox \newbox\gaxbox
\setbox\gaxbox=\hbox{\raise.5ex\hbox{$>$}\llap
     {\lower.5ex\hbox{$\sim$}}}\ht1=\grdimen\dp1=0pt
\setbox\laxbox=\hbox{\raise.5ex\hbox{$<$}\llap
     {\lower.5ex\hbox{$\sim$}}}\ht2=\grdimen\dp2=0pt
\def\gax{\mathrel{\copy\gaxbox}}
\def\lax{\mathrel{\copy\laxbox}}

\begin{document}
\title{Five Years in the Life of Cygnus X-1:\\ BATSE Long-Term Monitoring}

\author{W. S. Paciesas$^{*\sharp}$. C. R. Robinson$^{\flat\sharp}$, 
M. L. McCollough$^{\flat\sharp}$,\\ S. N. Zhang$^{\flat\sharp}$, 
B. A. Harmon$^\sharp$ and C. A. Wilson$^\sharp$}
\address{$^*$University of Alabama in Huntsville, AL 35899\\
$^\flat$Universities Space Research Association\\
$^\sharp$NASA/Marshall Space Flight Center, Huntsville, AL 35812}

\maketitle

\begin{abstract}
 The hard X-ray emission from Cygnus X-1 has been monitored
continually by BATSE since the launch of CGRO in April 1991. We present the
hard X-ray intensity and spectral history of the source covering a period of
more than five years. Power spectral analysis shows a significant peak at the
binary orbital period. The 20--100 \kev\ orbital light curve is roughly
sinusoidal with a minimum near superior conjunction of the X-ray source and an
rms modulation fraction of approximately 1.7\%. No longer-term
periodicities are evident in the power spectrum. We compare our results with
other observations and discuss the implications for models of the source
geometry.
\end{abstract}

\section*{Introduction}
Cyg X-1, the prototypical galactic black hole, is one of the most intensively
studied objects in the X-ray sky. Nevertheless, our understanding of its
detailed nature has been slow to evolve, due in large part to the strong but
rather chaotic variability of its X-ray emission. Long-term
monitoring in soft X-rays has been performed by several instruments, including
the all sky monitors on Ariel 5 \cite{Holt}, Vela 5B \cite{Priedh83},
and Ginga \cite{Kitamoto}. These have identified two main
periodic components in Cyg X-1, the 5.6 day binary orbital period and a less
well defined period of $\sim$300 days \cite{Priedh83}, the cause of
which is speculative.

The now well-known soft (``high'') and hard (``low'') emission states were also
identified from soft X-ray data. Such states are now known to be common
features of black holes, although not all black hole candidate systems have
shown both states (e.g., GS 2023+33, LMC X-1). 

With CGRO/BATSE we have now accumulated more than 5.5 years of monitoring of
Cyg X-1 in hard X-rays. Moreover, since the launch of RXTE we now have
wide band long-term monitoring of Cyg X-1. This enabled us
to obtain the most comprehensive observations yet of a complete soft state
episode \cite{Zhang}. We report
here exclusively on the BATSE long-term monitoring, including a broad overview
of the intensity and spectral variability and our search for periodic or
quasi-periodic components.

\section*{OBSERVATIONS}

Data obtained between 21 Apr 1991 and 24 Sep 1996 (TJD 8367--10350) were
processed using the standard BATSE Earth occultation software. Fluxes in the
energy range 20--100 \kev\ were calculated by fitting standard spectral models
to the 16-channel count spectra, either from individual occultation steps or
summed over one day. Two models were used: a single power-law and an optically
thin thermal bremsstrahlung (OTTB). In general, our
conclusions do not depend significantly on the choice of model spectrum.

\begin{figure}[ht]
\centerline{
\epsfig{file=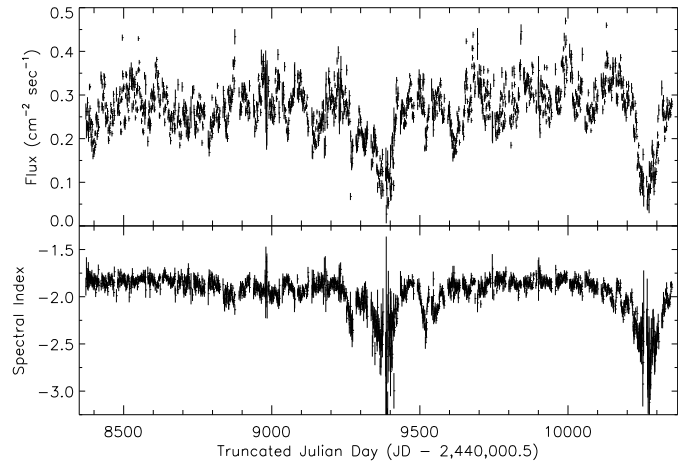,%
bbllx=62bp,bblly=54bp,bburx=528bp,bbury=734bp,height=3.5in,clip=,angle=90}}
\vspace{10pt}
\caption{Long-term hard X-ray intensity and spectral history of Cyg
X-1 from BATSE observations. The upper panel shows the integrated photon flux
in the 20--100 \kev\ band derived by fitting a power-law model to one-day
average count spectra. The lower panel shows the corresponding photon number
spectral index.}
\label{fig1}
\end{figure}

Figure~\ref{fig1} shows the long-term intensity and spectral history of Cyg X-1
in the 20--100 \kev\
energy range using one-day integrations. The two previously known
soft state episodes are clearly visible around TJDs 9350--9410 and
10220--10300. During the remaining time, the source intensity in the hard state
fluctuated rather randomly, staying mostly between 0.2 and 0.35 ph cm$^{-2}$
s$^{-1}$. The hard state spectral index remained relatively steady around a
value near $-1.85$, although extended periods with slightly softer spectra
occur, e.g., TJD 8950--9030. Flares above the 0.35 ph cm$^{-2}$ s$^{-1}$ level
typically last only a few days and show no spectral differentiation, whereas
dips below the 0.2 ph cm$^{-2}$ s$^{-1}$ level typically last a week or so, and
may or may not show spectral softening (cf. the intensity dips around TJDs 9510
and 9610).

\begin{figure}[t]
\begin{minipage}[t]{2.7in}
\mbox{}\\
\epsfig{file=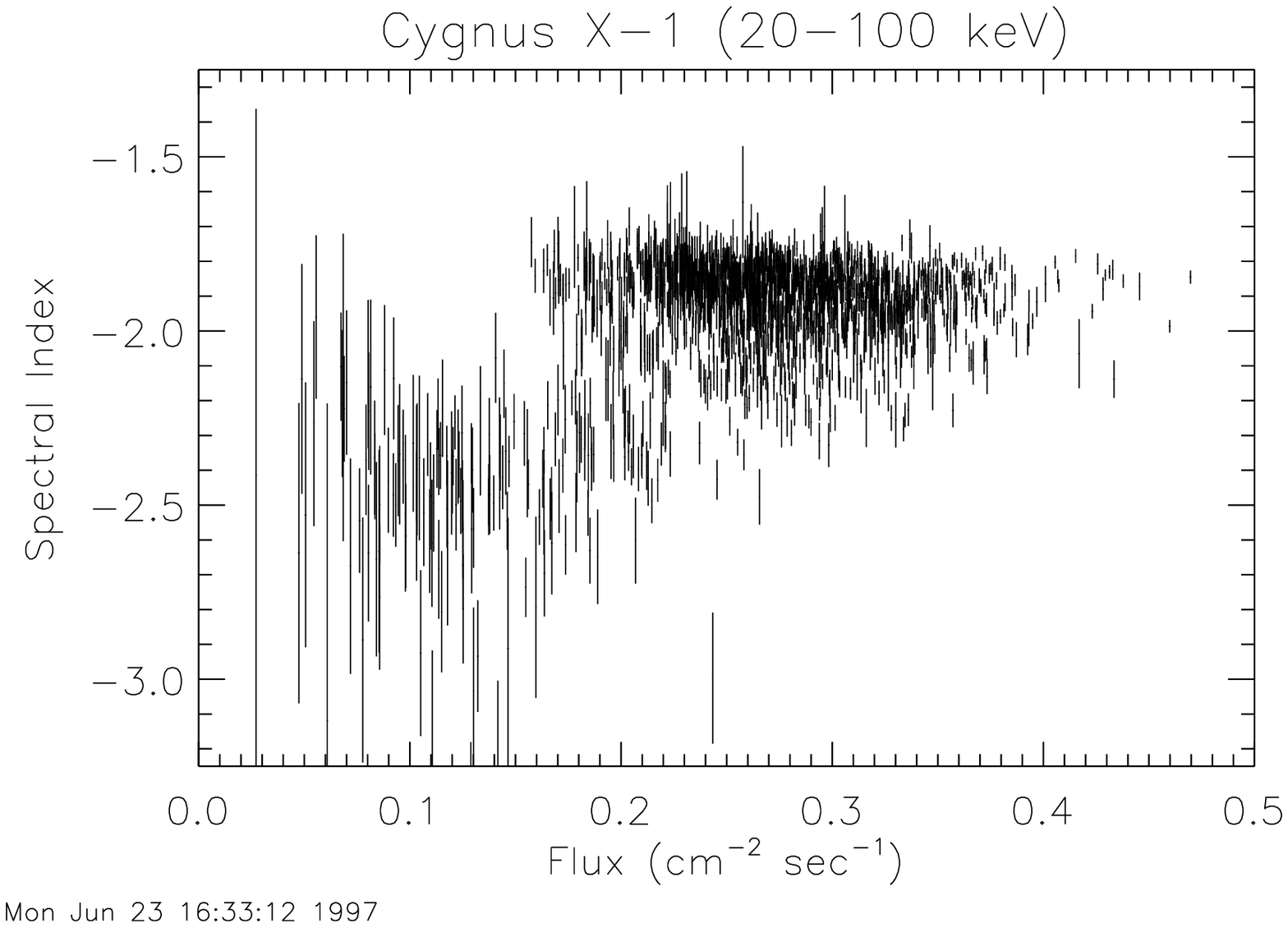,%
bbllx=57bp,bblly=375bp,bburx=548bp,bbury=690bp,height=1.7in,clip=}
\vspace{10pt}
\caption{\sloppy Correlation plot of spectral index vs.\ photon number flux
using the same data as in Figure~\ref{fig1}. For clarity, the flux errors are
not shown.}
\label{fig2}
\end{minipage}
\hfill
\begin{minipage}[t]{2.7in}
\mbox{}\\
\epsfig{file=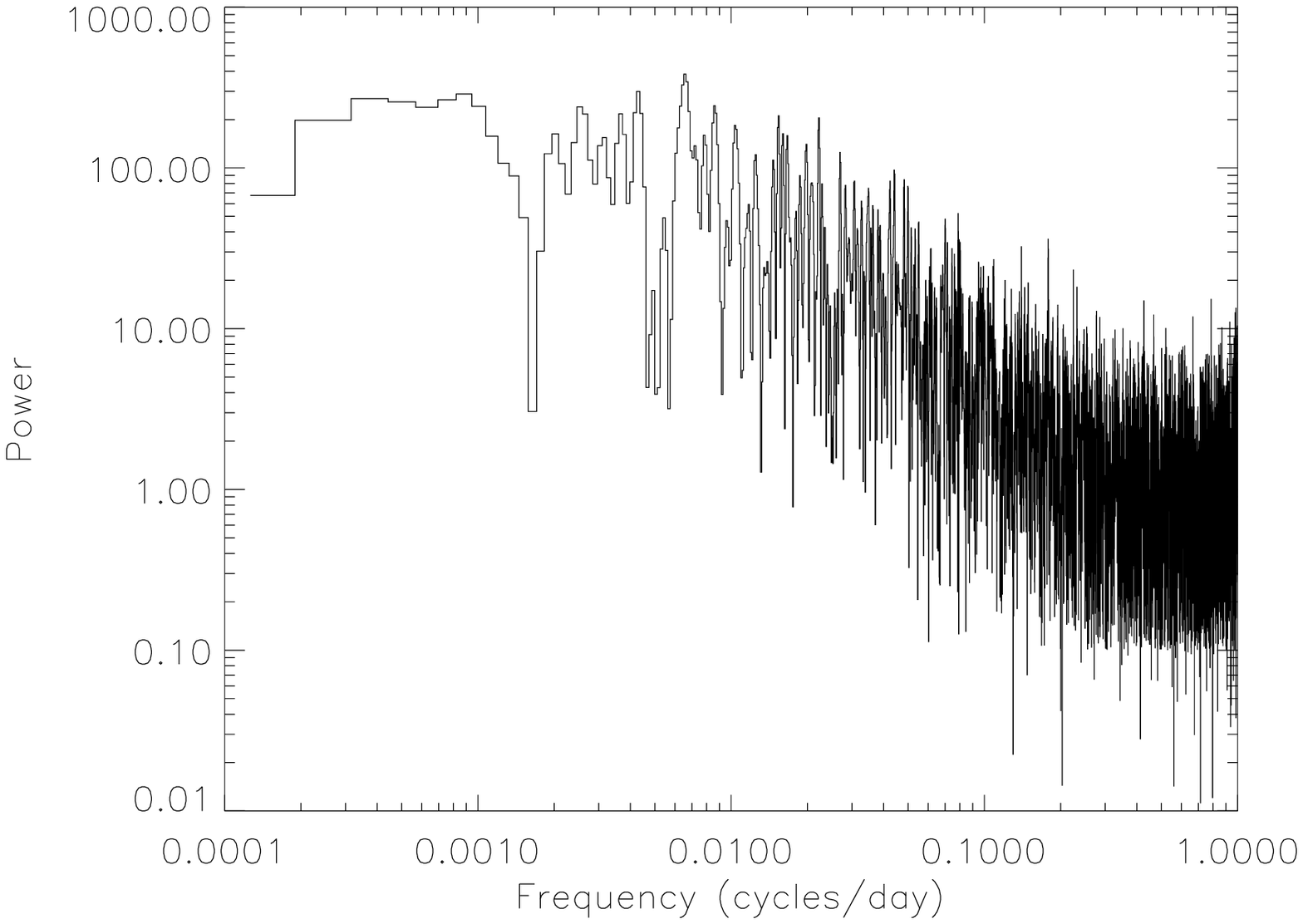,%
bbllx=63bp,bblly=360bp,bburx=532bp,bbury=700bp,height=1.9in,clip=}
\vspace{10pt}
\caption{The power density spectrum of Cyg X-1
hard X-ray flux, computed from single occultation step measurements
using the Scargle algorithm \protect\cite{Scargle} for unevenly sampled data.}
\label{fig3}
\end{minipage}
\end{figure}

Figure~\ref{fig2} shows a plot of the spectral index vs.\ flux for the same
data set. The predominance of the $-1.85$ spectral index over a wide range of
intensities is obvious, as is the trend toward softer spectral indices at low
intensity. However, the hard state can persist down to intensities as low as
0.15 ph cm$^{-2}$ s$^{-1}$ and spectra as soft as $\alpha\simeq -2.2$ can be
present at essentially any intensity. Spectra with $\alpha\lax -2.2$ are mainly
confined to flux levels below $\sim$0.15 ph cm$^{-2}$ s$^{-1}$, which appear to
occur only during the low energy ``high'' states. We do not see well-defined
intensity/spectral states corresponding to the three-state classification
scheme outlined by Ling et al.\ \cite{Ling83}.

To search for periodic and quasi-periodic signals, we used fluxes deconvolved
in a similar manner, but at the resolution of individual occultation steps. We
present results using the OTTB model, which produced slightly more robust fits.
The unevenly sampled power density spectrum (PDS) 
(Figure~\ref{fig3}) is rather flat at low
frequencies, falls off roughly as $1/f$ above $\sim$0.005 cycle/day (200 day
period), and reaches the Poisson noise level at $f\simeq 0.4$. To estimate the
significance of peaks in the data, we first fit the data with a combination of
a constant power $Z_0$ at low frequencies and a power-law $a_0 f^{a_1}$ above a
break frequency $f_{\rm c}$. The resulting parameters were  $Z_0=148$,
$a_0=0.333$, $a_1=-1.117$, and $f_{\rm c}=0.00425$. After dividing by the red
noise fit, the maximum of the power spectrum is at $f=0.178606$, consistent
with the binary orbit $f=0.178580$ \cite{Gies}. 
Treating this as an {\it a priori}
interesting frequency, the probability of a chance fluctuation is $1.4\times
10^{-7}$. If we ignore the {\it a priori} argument and scale by the number of
independent trial frequencies, the chance probability is
$5.4\times 10^{-5}$. Our data show no evidence for a peak around
$f=0.0033$ (300 day period); however, our sensitivity below $\sim$0.01 Hz is
limited by the red noise.

\begin{figure}[ht]
\centerline{\begin{minipage}{3.2in}
\epsfig{file=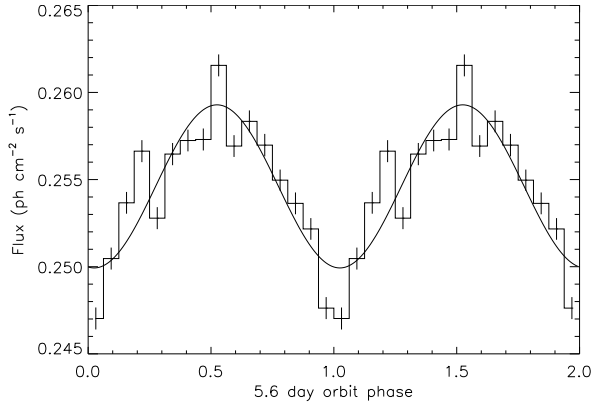,%
bbllx=60bp,bblly=367bp,bburx=545bp,bbury=695bp,height=2.1in,clip=}
\end{minipage}
\hfill
\begin{minipage}{2.2in}
\caption{\sloppy Cyg X-1 single occultation step data folded at the 5.6 day
binary orbital period. Two cycles are shown for clarity. The best-fitting sine
function is superimposed. Error bars represent the statistical error in the
mean for each phase bin. Phase 0 corresponds to the time when the supergiant
companion is closest to our line of sight.\protect\label{fig5}}
\end{minipage}}
\end{figure}

Figure~\ref{fig5} shows the data folded at the orbital period. The modulation
is roughly sinusoidal, with a minimum near phase 0 (supergiant companion
nearest the observer). The best-fitting sine function has a minimum at phase
$0.025\pm 0.008$ and peak-to-peak amplitude $0.0094\pm 0.0004$ ph cm$^{-2}$
s$^{-1}$ (statistical errors only), which corresponds to 3.8\% of the average
intensity. The rms scatter about the mean is $\sim1.7$\%.

\section*{Discussion}

Detection of the 5.6 day orbital variation in hard X-rays was first reported by
Ling et al.\ \cite{Ling90}, who found a peak-to-peak amplitude of $\sim$6\% in
50--140 \kev\ using $\sim$120 days of data from the HEAO-3 gamma ray
spectrometer. Priedhorsky et al.\ \cite{Priedh95} reported a marginal detection
of $10\pm 4$\% in 17--33 \kev\ from $\sim$70 days of observations with
WATCH/Eureca. Phlips et al.\ \cite{Phlips}, using $\sim$120 days of CGRO/OSSE
data, did not find a significant variation in 60--140 \kev, with an upper limit
to the rms fraction of 5\%. Our result is consistent with the OSSE upper limit,
and marginally consistent with the earlier observations.

Since the low energies show obvious absorption dips near phase 0
\cite{Holt,Mason,Murdin}, it is natural to consider absorption as responsible
for the variation we see. The decrease in our data centered roughly around
phase 0 cannot be due to absorption by cold matter because the column density
required would cause a total eclipse of the soft X-rays, which is not observed.
Electron scattering by highly ionized material 
would require a maximum column density 
$\sim 6 \times 10^{24}$ cm$^{-2}$. If the material is in a stellar wind,
the nearly sinusoidal shape we observe implies that this material is spread
over a large portion of the orbit. This is inconsistent with the much lower
column density ($\simeq 3\times 10^{23}$ cm$^{-2}$) estimated for such a wind
from 9--12 \kev\ data \cite{Priedh95}.

An alternative possibility is a variable reflection
component from the accretion disk or the companion star.
Done et al.\ \cite[also see ref.\
\protect\cite{Haardt}]{Done} showed that the Cyg X-1 spectrum can be fit with a
model involving reflection from an ionized accretion disk, similar to models
for active galactic nuclei. In these fits, the reflection
component represents $\sim$30\% of the flux in the 20--100 \kev\ range, so that
our results could be explained by a variation of 5--10\% in reflectivity as
a function of phase.

\section*{Summary}

BATSE has observed Cyg X-1 continually for more than 5.5 years. The hard X-ray
light curve is dominated by red noise that has a flat power spectrum below a
frequency of $\sim$0.004 cycle/day (periods $\simeq$ 250 days) and falls off
roughly as 1/$f$ at higher frequencies. Periodic variability is detected at the
binary orbital period, with an rms modulation of $\sim$1.7\% and a minimum flux
at the time of superior conjunction of the supergiant companion (phase 0).
We find no evidence for the previously
reported long-term period of $\sim$300 days.

The 20--100 \kev\ spectrum of Cyg X-1 appears to have a spectral hardness limit
around a power-law index $\alpha\simeq -1.8$. BATSE has observed such a
spectrum over a flux range of at least a factor of three, from $\sim$0.15 to
$\gax$0.45 ph cm$^{-2}$ s$^{-1}$. However, softer spectra can be present at any
observed flux level. Below $\sim$0.15 ph cm$^{-2}$ s$^{-1}$, only soft spectra
($\alpha\lax -2.25$) associated with the soft (high) X-ray state have so far
been seen.

\end{document}